\documentclass[aps,pra,preprint,superscriptaddress,longbibliography]{revtex4-1}

\usepackage{amsmath,amsthm,amssymb}
\usepackage{xcolor}
\usepackage{graphicx}
\usepackage{bbold}
\usepackage{textcomp} 
\usepackage{siunitx}
\usepackage{ifthen}

\renewcommand{\vec}[2][\empty]{
  \ifthenelse{\equal{#1}{\empty}}
  {\boldsymbol{#2}}
  {\boldsymbol{#2}_\text{#1}}
}

\newcommand{\vnabla}{\vec{\nabla}}

\def\tausf{\tau_\text{sf}}

\def\Ms{M_\text{s}}

\newcommand{\updownarrows}{\uparrow\mathrel{\mspace{-1mu}}\downarrow}
\newcommand{\downuparrows}{\downarrow\mathrel{\mspace{-1mu}}\uparrow}
\renewcommand{\upuparrows}{\uparrow\uparrow}
\renewcommand{\downdownarrows}{\downarrow\downarrow}

\begin{document}

\title{
  Back hopping in spin-transfer-torque devices, possible origin and counter measures
}

\author{Claas Abert}
\email[]{claas.abert@univie.ac.at}
\affiliation{Christian Doppler Laboratory of Advanced Magnetic Sensing and Materials, Faculty of Physics, University of Vienna, Austria}

\author{Hossein Sepehri-Amin}
\affiliation{National Institute for Materials Science, Tsukuba 305-0047, Japan}

\author{Florian Bruckner}
\affiliation{Christian Doppler Laboratory of Advanced Magnetic Sensing and Materials, Faculty of Physics, University of Vienna, Austria}

\author{Christoph Vogler}
\affiliation{Faculty of Physics, University of Vienna, Austria}

\author{Masamitsu Hayashi}
\affiliation{National Institute for Materials Science, Tsukuba 305-0047, Japan}
\affiliation{Department of Physics, The University of Tokyo, Bunkyo, Tokyo 113-0033, Japan}

\author{Dieter Suess}
\affiliation{Christian Doppler Laboratory of Advanced Magnetic Sensing and Materials, Faculty of Physics, University of Vienna, Austria}

\date{\today}

\begin{abstract}
  The effect of undesirable high-frequency free-layer switching in magnetic multilayer systems, referred to as back hopping, is investigated by means of the spin-diffusion model.
  A possible origin of the back-hopping effect is found to be the destabilization of the pinned layer which leads to perpetual switching of both layers.
  The influence of different material parameters on the critical switching currents for the free and pinned layer is obtained by micromagnetic simulations.
  It is found that the choice of a free-layer material with low polarization $\beta$ and saturation magnetization $M_\text{s}$, and a pinned-layer material with high $\beta$ and $M_\text{s}$ leads to a low free-layer critical current and a high pinned-layer critical current and hence reduces the likelihood of back hopping.
  While back hopping was observed in various types of devices, there are only few experiments that exhibit this effect in perpendicularly magnetized systems.
  However, our simulations suggest, that this is likely to change due to loss of pinned-layer anisotropy when decreasing device sizes.
\end{abstract}

\pacs{}

\maketitle
  
\section{Introduction}
Spin-transfer torque (STT) in magnetic multilayers has gained a lot of interest in recent years due to possible applications in novel storage devices.
A prominent candidate for such an STT magnetic random access memory (MRAM) is a trilayer system consisting of two magnetic layers separated by a nonmagnetic layer \cite{huai2008spin,lin200945nm,khvalkovskiy2013basic}.
If an electric current passes this system, one of the magnetic layers acts as a spin polarizer.
The other layer is subject to the spin torque exerted by the spin polarized electrons.
Depending on the sign of the electric current, the magnetization of this free layer can be switched in either direction.
Since the spin-torque coupling is bidirectional, the spin polarizing layer, also referred to as pinned layer, is usually constructed to be very stiff to prevent switching.

The spacer layer between the magnetic layers can either be a conductor or insulator.
In case of an insulator, the spin polarized electrons must tunnel through the spacer in order to exert a torque on the free-layer magnetization.
The magnetization in the magnetic layers can be either in-plane or out-of-plane.
In the case of in-plane magnetization, the pinned layer is mainly stabilized by its thickness which leads to a high shape anisotropy.
In the case of out-of-plane magnetization, the pinned layer is a magnetic multilayer system with high uniaxial anistropy.

It was observed in different in-plane devices that the free-layer magnetization might be unstable after switching.
This back-hopping effect happens after overcoming the critical switching current and results in fast switching of the free layer \cite{min2009back,pufall2004large,urazhdin2003current,devolder2016time}.
Different explanations for this effect were proposed \cite{sun2009high}.
One possible explanation is the destabilization of the pinned layer that causes the perpetual switching of both the free layer and the pinned layer \cite{hou2011dynamics}.
In this work we investigate this behaviour by means of micromagnetic simulations.

\section{Model}\label{sec:model}
According to the micromagnetic model, the magnetization dynamics are governed by the Landau-Lifshitz-Gilbert equation (LLG)
\begin{equation}
  \frac{\partial \vec{m}}{\partial t} =
  - \gamma \vec{m} \times \left( \vec[eff]{h} + \frac{J}{\hbar \gamma \Ms} \vec{s} \right)
  + \alpha \vec{m} \times \frac{\partial \vec{m}}{\partial t},
  \label{eq:llg}
\end{equation}
where $\vec{m}$ is the normalized magnetization, $\gamma$ is the gyromagnetic ratio, $\alpha$ is the Gilbert damping, and $\vec[heff]{h}$ is the effective field that usually contains the demagnetization field, the exchange field, as well as other contributions depending on the problem setting.
The effective field is complemented by a contribution from the spin accumulation $\vec{s}$ with $M_\text{s}$ being the saturation magnetization and $J$ being the exchange integral of the itinerant electrons and the magnetization.
The spin accumulation $\vec{s}$ describes the deviation of the magnetic moment carried by the conducting electrons with applied current from the equilibrium situation without current.
According to the spin-diffusion model \cite{zhang2002mechanisms} $\vec{s}$ is defined by the equation of motion
\begin{equation}
  \frac{\partial \vec{s}}{\partial t} =
  - \vnabla \cdot \vec[s]{j}
  - \frac{\vec{s}}{\tausf}
  - J \frac{\vec{s} \times \vec{m}}{\hbar}
  \label{eq:spin_accumulation}
\end{equation}
where $\tausf$ is the spin-flip relaxation time and $\vec[s]{j}$ is the matrix-valued spin current defined by
\begin{equation}
  \vec[s]{j} =
  \beta \frac{\mu_\text{B}}{e} \vec{m} \otimes \vec[e]{j}
  - 2 D_0 \left[
    \vnabla\vec{s}
    - \beta \beta' \vec{m} \otimes ( (\vnabla\vec{s})^T \vec{m})
  \right],
  \label{eq:spin_current_simple}
\end{equation}
where $D_0$ is the diffusion constant and $\beta$ and $\beta'$ are dimensionless polarization parameters.
While $\beta$ is a measure of the capability of a material to polarize itinerant electrons, $\beta'$ is measure for the sensitivity of the electric resistivity to the angle between magnetization and polarization of itinerant electrons.

Instead of performing a time integration of the spin accumulation $\vec{s}$ along with the magnetization $\vec{m}$, we assume $\vec{s}$ to be in equilibrium at all times, i.e. we compute $\vec{s}(\vec{m})$ by $\partial \vec{s}(\vec{m})/\partial t = 0$.
Since the spin accumulation relaxes orders of magnitude faster than the magnetization \cite{zhang2004roles}, this assumption does not have a considerable effect on the magnetization dynamics.
The LLG coupled to the spin diffusion model is solved numerically with the finite-element method.
The numerical solution of this system is described in detail elsewhere \cite{ruggeri2015coupling}.

\begin{figure}
  \includegraphics{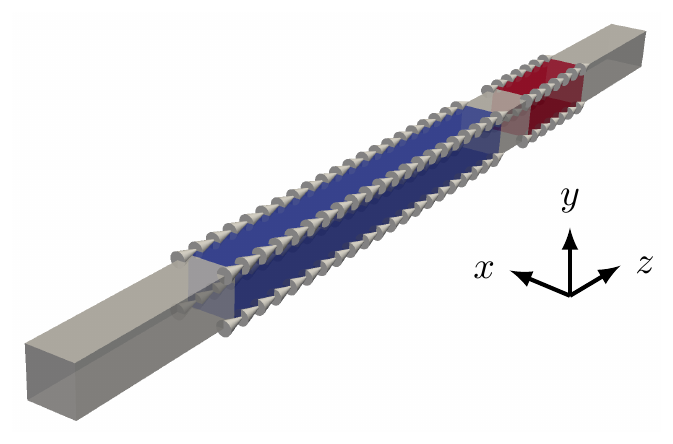}
  \caption{
    The quasi one-dimensional model with pinned layer (blue), free layer (red), spacer and leads (grey).
  }
  \label{fig:model}
\end{figure}
For the investigation of the back-hopping effect we consider a system with a pinned-layer thickness of $\SI{10}{nm}$, a spacer-layer thickness of $\SI{1.5}{nm}$ and a free-layer thickness of $\SI{3}{nm}$.
Additionally, the trilayer is sandwiched between two nonmagnetic leads.
These leads are simulated as layers with a thickness of $\SI{4}{nm}$.
However, due to the use of effective material parameters, the simulation results are similar to those of inifinite leads \cite{abert2016field}.
The model is quasi one-dimensional, i.e. the cross section of the simulated device is chosen to be a square with dimensions \SI{1x1}{nm}, see Fig.~\ref{fig:model}.
Since the size of a typical STT MRAM device is considered to be below the single-domain limit, the choice of lateral dimension is considered to be a valid assumption.

We consider out-of-plane magnetized systems in this work.
In these systems the pinned layer is mainly stabilized by a high uniaxial anisotropy.
The magnetic material parameters for the pinned layer are chosen as $\Ms = \SI{1.4}{T}$, $K_1 = \SI{e6}{J/m^3}$, and $A = \SI{e-11}{J/m}$ which is typical for FePt.
For the free layer we chose material parameters $\Ms = \SI{1.357}{T}$ $K_1 = \SI{2e5}{J/m^3}$, and $A = \SI{3e-11}{J/m}$.
The remaining material parameters for both magnetic layers are chosen as $\alpha = 0.02$, $\beta = \beta' = 0.8$, $D_0 = \SI{e-3}{m^2/s}$, $\tausf = \SI{5e-14}{s}$, and $J = \SI{6e-20}{J}$.
The spacer-layer as well as the leads are simulated with material parameters similar to Ag, namely $D_0 = \SI{5e-3}{m^2/s}$ and $\tausf = \SI{e-12}{s}$.
The coordinate system is chosen such that the $z$-axis points out of plane.
In our simplified model we only consider the exchange field and the anisotropy field as effective-field contributions.
While the demagnetization field certainly has an impact on real systems by introducing shape anisotropy and interlayer coupling, it is not considered to have a qualitative impact on the results presented in this work.
Moreover, omitting the demagnetization field justifies the use of the quasi one-dimensional model, since the simulation results do not depend on the lateral dimension in this case.

\section{Back Hopping}
\begin{figure}
  \centering
  \includegraphics{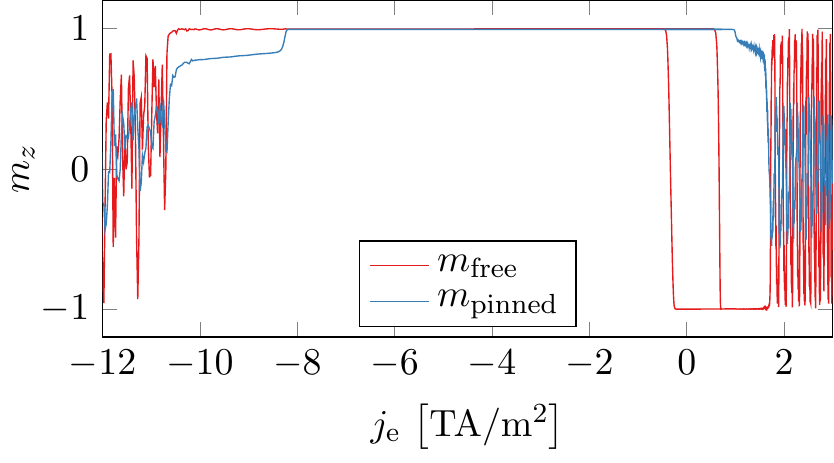}
  \caption{
    Full current hysteresis loop of the presented STT MRAM structure.
  }
  \label{fig:hysteresis}
\end{figure}
Fig.~\ref{fig:hysteresis} shows the current hysteresis loop for the model introduced above.
The effect of back hopping can be observed on both branches of the hysteresis loop.
However, on the positive current branch, the back hopping happens at much lower currents.
The initial situation is a parallel configuration of the pinned and free layer $m_{\text{free},z} = m_{\text{pinned},z} = 1$.
A positive current means that electrons are flowing from the free layer to the pinned layer.
In this situation, the spin torque in the free layer is generated indirectly by electrons scattered from the pinned-layer--spacer-layer interface.
After the free layer switches at a current density of $j_\text{e} = \SI{7e11}{A/m^2}$, the back hopping can be observed at a current density of $j_\text{e} = \SI{1.7e12}{A/m^2}$ and higher.
It is clear from Fig.~\ref{fig:hysteresis} that the back hopping of the free layer is initiated by switching of the pinned layer.

\begin{figure}
  \includegraphics{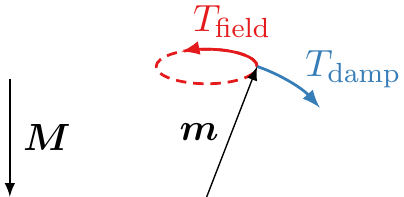}
  \caption{
    Field-like and damping-like torque caused by a spin polarized current with polarization $\vec{M}$ acting on the magnetization $\vec{m}$.
  }
  \label{fig:torques}
\end{figure}
In order to understand the perpetual switching of both layers, the spin torque acting on both layers has to be investigated in detail.
The overall spin torque is usually split up into a field-like and a damping-like torque on the basis of the spin polarization $\vec{M}$ of the electric current.
The damping-like torque, that is held mainly responsible for the switching, is given by
\begin{equation}
  \vec{T}_\text{damp} = \vec{m} \times (\vec{m} \times \vec{M}).
\end{equation}
Since the spin accumulation enters the LLG via a torque term of the form $\vec{m} \times \vec{s}$, the strength of the damping-like torque with respect to a given reference polarization $\vec{M}$ can be computed from the spin accumulation $\vec{s}$ by projection on $\vec{m} \times \vec{M}$.
In contrast to the field-like torque that induces precessional motion of the free-layer magnetization, the damping-like torque leads to a direct relaxation of the free-layer magnetization either parallel or antiparallel to the pinned layer, see Fig.~\ref{fig:torques}.

We consider a system with perpendicular uniaxial anisotropy in both the free layer and the pinned layer.
While the anisotropy axis in the free layer is perfectly aligned with the $z$-axis, the anisotropy axis of the pinned layer is slightly tilted in $y$-direction in order to break symmetry and avoid metastable states.
This leads to an equilibrium magnetization of $\vec[free]{m} = \pm (0,0,1)$ in the free layer and $\vec[pinned]{m} = \pm (0,\epsilon,1)$ in the pinned layer, where $\epsilon \ll 1$ reflects the tilted anisotropy axis in the pinned layer.
When computing the spin torque in the free layer, the pinned layer is considered to act as a spin polarizer and thus the reference polarization is defined by $\vec{M} = \vec[pinned]{m}$.
The strength of the damping-like torque in the free layer is then obtained by projection of the accumulation $\vec{s}$ on $\vec[free]{m} \times \vec[pinned]{m}$.
The damping-like torque in the pinned layer is generated by the spin polarized currents coming from the free layer and thus $\vec{M} = \vec[free]{m}$.
Accordingly, the strength of the damping-like torque in the pinned layer is given by the projection of $\vec{s}$ on $\vec[pinned]{m} \times \vec[free]{m}$.
This means that for all equilibrium magnetization configurations in the model system ($\upuparrows$, $\downdownarrows$, $\updownarrows$, $\downuparrows$) the strength of the damping-like torque in both the free and the pinned layer is either $\epsilon s_x$ or $-\epsilon s_x$ depending on the exact configuration.
Namely for parallel configurations ($\upuparrows / \downdownarrows$), a positive $s_x$ results in a positive damping-like torque in the free layer and a negative damping-like torque in the pinned layer.
On the other hand, for antiparallel configurations ($\updownarrows / \downuparrows$), a positive $s_x$ results in a negative damping-like torque in the free layer and a positive damping-like torque in the pinned layer.
A positive damping-like torque leads to a parallel alignment of the magnetization with the reference polarization $\vec{M}$ while a negative torque leads to an antiparallel alignment with the reference polarization.

\begin{figure}
  \includegraphics{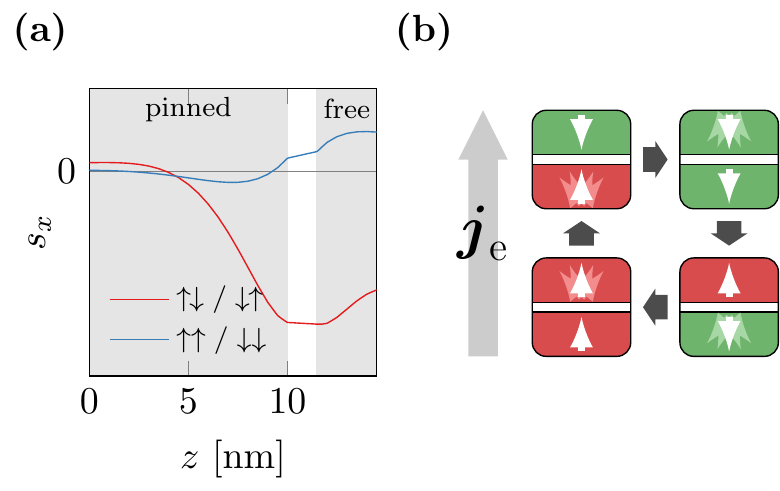}
  \caption{
    Spin-accumulation and switching process of a magnetic trilayer in the back-hopping regime.
    (a) $x$-component of the spin accumulation for parallel and antiparallel magnetization configuration with pinned layer slightly tilted in $y$-direction.
    (b) Cyclic switching process of free layer (top) and pinned layer (bottom).
  }
  \label{fig:hopping}
\end{figure}
Fig.~\ref{fig:hopping}(a) shows the $x$-component of the spin accumulation $\vec{s}$ for a current $\vec[e]{j}$ in positive $z$-direction and the different magnetization configurations introduced above.
It is worth noting that the spin accumulation $s_x$ only differs for parallel/antiparallel configurations.
Considering these results, the switching process of the two magnetic layers in the back-hopping regime can be understood as follows.
We assume both magnetic layers to be magnetized in positive $z$-direction in the beginning, see the left bottom picture of Fig.~\ref{fig:hopping}(b).
For this configuration, $s_x$ is positive in the free layer, which corresponds to a negative damping-like torque, and $s_x$ is negative in the pinned layer, which corresponds to a positive damping-like torque.
This results in destabilization of the free layer and stabilization of the pinned layer respectively and hence leads to switching of the free layer.
For the resulting antiparallel configuration $s_x$ is negative in both the free layer and the pinned layer.
Hence, the free layer is subject to negative damping-like torque while the pinned layer is subject to positive damping-like torque.
As a result, the pinned layer switches.
This behavior leads to a cyclic process depicted in Fig.~\ref{fig:hopping}(b), which explains the perpetual switching of both layers.
Note, that in every step of this cycle the spin torque stabilizes one of the two layers while it destabilizes the other one.

\begin{figure}
  \centering
  \includegraphics{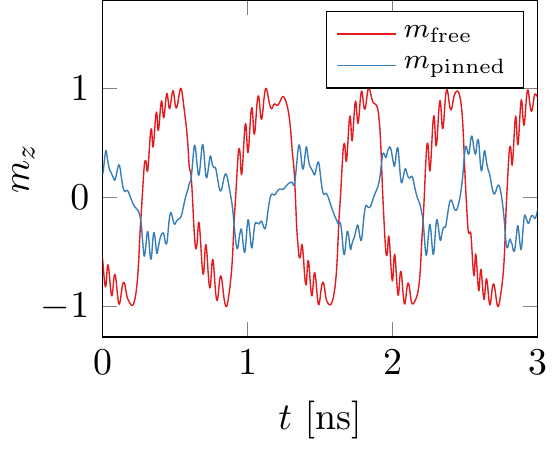}
  \caption{
    Time evolution of the magnetization in the back-hopping regime.
  }
  \label{fig:hysteresis_detail}
\end{figure}
The above explanation of the back-hopping effect assumes that the distinctive steps of the cycle happen one after another.
However, dynamical simulations suggest that the switching of the two layers happens in a more dynamic fashion, see Fig.~\ref{fig:hysteresis_detail}.
Especially it should be noted that the pinned layer never reaches saturation during the oscillation process.

While the model system is perpendicularly magnetized, the same cyclic process can also be reproduced in in-plane magnetized multilayer structures.
In fact, in-plane systems are expected to be more prone to back hopping, since the pinned layer in such systems is only stabilized by shape anisotropy.
Perpendicular systems, on the other hand, exploit anisotropies of magnetic multilayers to stabilize the pinned layer, which enables a better control of the anisotropy strength of the pinned layer.
This consideration is supported by experimental data.
While different experimental studies demonstrate back hopping for in-plane systems \cite{pufall2004large, urazhdin2003current} as well as perpendicular systems \cite{devolder2016time,kim2016experimental}, the effect has been considered to be a minor problem for perpendicular systems \cite{nowak2011demonstration}.
However, with devices shrinking in size \cite{yuasa2013future}, it becomes more challenging to stabilize the pinned layer \cite{sato2013comprehensive}.
Hence, back hopping is expected to become a serious issue for perpendicular systems, too.
Note, that the diffusion model, which is used thoughout this work, applies to metallic junctions, while modern perpendicular MRAM devices are usually magnetic tunnel junction (MTJ).
However, the back-hopping effect was also observed in MTJs \cite{oh2009bias} and the general mechanism of the hopping process is expected to be the same for ballistic transport as for diffusive transport.

\section{Counter measures}
In order to design a reliable STT MRAM device, it is important to prevent back hopping since it puts the device in a nonpredictable state.
Hence, the material parameters of the different layers should be chosen such that the critical current for free-layer switching is well below the critical current for pinned-layer switching.
In the following we present the critical currents for both free layer and pinned layer for a perpendicular system with a parallel initial magnetization configuration.
This means, that the critical current for the free layer indicates the switch from parallel to antiparallel configuration and the critical current for the pinned layer indicates the switch back to the parallel configuration.
If not stated differently, the geometry and material parameters of the system are the same as introduced previously.
The critical currents are obtained by linearly increasing the current density with a rate of $\SI{0.2e21}{A/m^2s}$ and determining the current at switching.

\begin{figure}
  \center
  \includegraphics{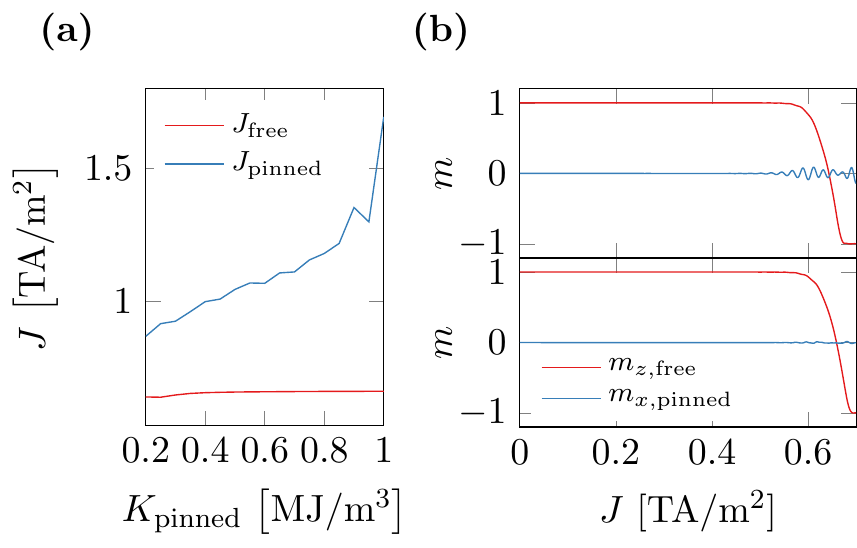}
  \caption{
    Free-layer switching for different pinned-layer anisotropies.
    (a) Critical current densities for switching of the free layer and pinned layer depending on the pinned-layer anisotropy constant $K_\text{pinned}$.
    (b) Switching process for linearly ramped current for $K_\text{pinned} = \SI{0.2}{MJ/m^3}$ (top) and $K_\text{pinned} = \SI{0.4}{MJ/m^3}$ (bottom).
  }
  \label{fig:vary_k}
\end{figure}
Figure~\ref{fig:vary_k}(a) shows the critical currents for different pinned-layer anisotropies.
It doesn't come as a surprise that the free-layer critical current is almost independent from $K_\text{pinned}$ while the fixed-layer critical current increases with increasing $K_\text{pinned}$.
However, it should be noted that the free-layer critical current shows a slight decrease of approximately 3\% for very small values of $K_\text{pinned}$.
This can be explained with the excitation of the pinned layer which assists the switching of the free layer, see Fig.~\ref{fig:vary_k}(b).

\begin{figure}
  \includegraphics{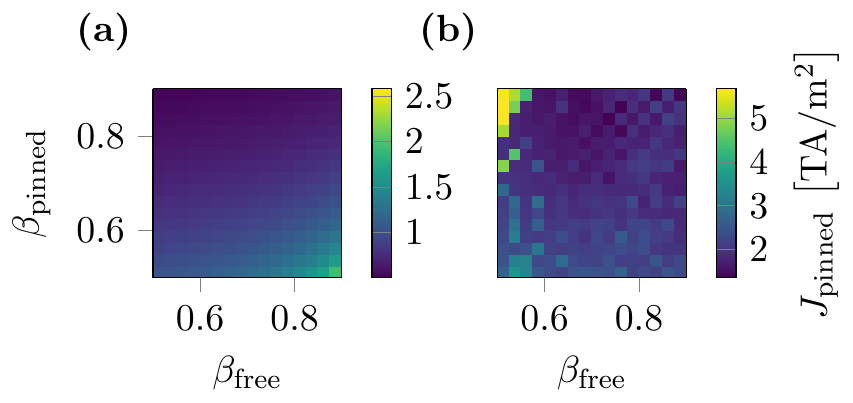}
  \caption{
    Critical current densities for various polarization parameter $\beta$ of free and pinned layer.
    (a) Critical current for free-layer switching.
    (b) Critical current for pinned-layer switching.
  }
  \label{fig:vary_beta}
\end{figure}
Another promising material parameter for critical-current manipulation is the polarization $\beta$ both in the free and the pinned layer.
Figure~\ref{fig:vary_beta} shows the critical currents for different values of $\beta_\text{pinned}$ and $\beta_\text{free}$.
Since $\beta$ is a measure for the ability of a material to polarize itinerant electrons, it is expected that a large $\beta_\text{pinned}$ will decrease the critical current for free-layer switching and a small $\beta_\text{free}$ will increase the critical current for pinned-layer switching as desired.
This behaviour is well reflected by the numerical experiments.
Moreover, the data clearly shows that a small $\beta_\text{free}$ decreases the critical current for free-layer switching as well as a large $\beta_\text{pinned}$ increases the critical current for pinned-layer switching.
This effect is not obvious when considering the switching of one layer to be initiated mainly by polarized electrons coming from the other layer.
However, a highly polarizing material does not only emit highly polarized electrons, but also strongly depolarizes incoming electrons with a different polarization which well explains this effect.
In conclusion, materials should be chosen to have a large $\beta_\text{pinned}$ and a small $\beta_\text{free}$ in order to avoid back hopping.

\begin{figure}
  \includegraphics{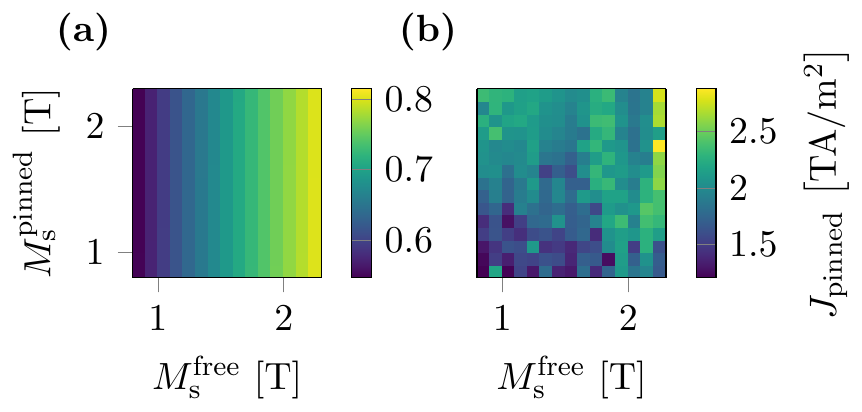}
  \caption{
    Critical current densities for various saturation magnetizations $M_\text{s}$ of free and pinned layer.
    (a) Critical current for free-layer switching.
    (b) Critical current for pinned-layer switching.
  }
  \label{fig:vary_ms}
\end{figure}
Another material parameter that is expected to influence the critical currents is the saturation magnetization $M_\text{s}$ of the individual layers.
Fig.~\ref{fig:vary_ms} shows the simulation results for varying $M_\text{s}^\text{pinned}$ and $M_\text{s}^\text{free}$.
The simulations show that the free-layer critical current $I_\text{free}$ depends on the free-layer saturation magnetization $M_\text{s}^\text{free}$ only, see Fig.~\ref{fig:vary_ms}(a).
The independence of $I_\text{free}$ on $M_\text{s}^\text{pinned}$ is well explained by the fact, that the solution of the spin accumulation \eqref{eq:spin_accumulation} does not depend on the saturation magnetization.
However, both the spin torque \eqref{eq:llg} and the anistropy field $H_\text{aniso} = 2 K m_z/\mu_0 M_\text{s}$ scale with $1/M_\text{s}$.
Since the critical current is a result from the competition of these two contributions, it is quite surprising that the simulated critical currents show a clear dependence on the free-layer saturation magnetization $M_\text{s}^\text{free}$.
The origin of this effect, which is also found in experiments \cite{yagami2004low}, is the dependence of the characteristic switching time on the saturation magnetization $M_\text{s}^\text{free}$.
While, strictly speaking, the critical current remains unchanged for different $M_\text{s}^\text{free}$, a lower $M_\text{s}^\text{free}$ leads to faster switching.
Since the critical current, as presented in Fig.~\ref{fig:vary_ms}, is determined by linearly increasing the current in time, low switching times directly lead to low critical currents.
The details of this effect will be discussed elsewhere.

Note, that a similar dependence should be found for the pinned-layer critical current.
However, while Fig.~\ref{fig:vary_ms}(b) shows the same trend of a larger critical current for larger $M_\text{s}^\text{pinned}$, the simulation results are very noisy compared to Fig.~\ref{fig:vary_ms}(a).
This noise can also be observed in Fig.~\ref{fig:vary_beta}(b).
The reason for the noise lies in the stiffness of the pinned layer.
After the free layer has switched, the spin accumulation leads to a stabilization of the free layer and a destabilization of the pinned layer.
Since the pinned layer is much stiffer than the free layer, large currents are required to push the pinned layer out of its equilibrium.
Also the pinned layer is not instantaneously switched, but slightly tilts and moves with a high frequency, see Fig.~\ref{fig:hysteresis}.
In this intermediate state, the dynamics of the pinned layer generates a dynamic spin accumulation that ultimately also excites the free layer.
Due to the complexity of this coupling, the critical switching current for the pinned layer is very sensitive to perturbations of the system, which leads to the observed noise in the simulation results.
This noise is also expected to occur in experiments where it might even be more significant due to thermal effects.

\section{Conclusion}
We have investigated the back-hopping effect in perpendicularly magnetized STT MRAM devices by means of the spin-diffusion model.
Undesired switching of the pinned layer has been found to be a possible origin for the back-hopping effect that leads to fast oscillations of the free-layer magnetization.
A possible solution to avoid the switching of the pinned layer is the increase of the pinned-layer anisotropy.
However, decreasing the size of MRAM devices, in order to increase the storage density, will lead to lower anisotropies and thus increase the chances for back hopping.

Our numerical studies suggest that a high polarization $\beta_\text{pinned}$ of the pinned layer and a low saturation magnetization $M_\text{s}^\text{free}$ in the free layer result in a low critical current for free-layer switching.
Similarly, a low $\beta_\text{free}$ and a high $M_\text{s}^\text{pinned}$ result in a high critical current for pinned layer switching, and thus back hopping.

\appendix
\section*{Acknowledgements}
CA would like to thank Kazuhiro Hono for making this collaborative work possible.
The financial support by
the Austrian Federal Ministry of Science, Research and Economy and the National Foundation for Research, Technology and Development
as well as
the Austrian Science Fund (FWF) under grant F4112 SFB ViCoM,
the Vienna Science and Technology Fund (WWTF) under grant MA14-44,
and MEXT Grant‐in‐Aids for Scientific Research (16H03853) of Japan
is gratefully acknowledged.

\end{document}